\documentclass[11pt,a4paper]{article}
\usepackage[cm]{fullpage}

\usepackage{graphicx}
\usepackage{epsfig,amssymb,graphicx,lscape,color,lineno,enumitem}
\usepackage{setspace}
\usepackage{amsmath}
\usepackage[usenames,dvipsnames]{xcolor}
\title{Predicting population extinction in lattice-based birth-death-movement processes}
\author{Stuart T. Johnston$^{1,2,*}$, Matthew J. Simpson$^{3}$, Edmund J. Crampin$^{1,2,4}$}

\date{}
\makeatletter
\newcommand{\vast}{\bBigg@{3}}
\newcommand{\Vast}{\bBigg@{4}}
\makeatother

\begin{document}
\maketitle
\begin{abstract}
The question of whether a population will persist or go extinct is of key interest throughout ecology and biology. Various mathematical techniques allow us to generate knowledge regarding individual behaviour, which can be analysed to obtain predictions about the ultimate survival or extinction of the population. A common model employed to describe population dynamics is the lattice-based random walk model with crowding (exclusion). This model can incorporate behaviour such as birth, death and movement, while including natural phenomena such as finite size effects. Performing sufficiently many realisations of the random walk model to extract representative population behaviour is computationally intensive. Therefore, continuum approximations of random walk models are routinely employed. However, standard continuum approximations are notoriously incapable of making accurate predictions about population extinction. Here, we develop a new continuum approximation, the \textit{state space diffusion approximation}, which explicitly accounts for population extinction. Predictions from our approximation faithfully capture the behaviour in the random walk model, and provides additional information compared to standard approximations. We examine the influence of the number of lattice sites and initial number of individuals on the long-term population behaviour, and demonstrate the reduction in computation time between the random walk model and our approximation.
\end{abstract}


\noindent $^1$ Systems Biology Laboratory, School of Mathematics and Statistics, and Department of Biomedical Engineering, University of Melbourne, Parkville, Victoria, Australia. \\
$^2$ ARC Centre of Excellence in Convergent Bio-Nano Science and Technology, Melbourne School of Engineering, University of Melbourne, Parkville, Victoria, Australia. \\
$^3$ School of Mathematical Sciences, Queensland University of Technology, Brisbane, Queensland, Australia. \\
$^4$ School of Medicine, Faculty of Medicine Dentistry and Health Sciences, University of Melbourne, Parkville, Victoria, Australia. \\
$^*$ e-mail: stuart.johnston@unimelb.edu.au

\newpage

\section{Introduction}

Determining whether the behaviour of individuals in a population results in the extinction or survival of the population is a key question in cell biology, population biology and ecology \cite{anderson1992,axelrod2006,berger1990,brightwell2018,cantrell1998,keeling2007,neufeld2017,saltz1995,thomas2004}. For example, tumour cells can undergo migration from the original tumour site to form metastases \cite{coghlin2010}. However, this relies on the successful colonisation of a new site by a small number of tumour cells. As such, the question of whether the new colony of cells will survive or go extinct is paramount \cite{korolev2014}. Similarly, in the field of drug development, determining whether a novel treatment results in the eradication of a target cell population is crucial \cite{schmidt2011}. In an ecological context, predicting whether the introduction of a threatened species to a particular area will be successful is critical for selecting suitable sites \cite{berger1990,saltz1995}. \\

Mathematical models provide a framework for incorporating knowledge about population behaviour such that the evolution of the population can be evaluated \cite{anderson1992,brightwell2018,chaplain2019,fadai2019,gerlee2013,gourley2003,johnson2019,neufeld2017,sarapata2014}. The most common mathematical models used to investigate population dynamics are \textit{continuum} models, such as ordinary differential equation (ODE) or partial differential equation (PDE) models. These approaches aim to predict the temporal and spatiotemporal distribution of the average density of individuals, respectively \cite{gatenby1996,gatenby2006,gerlee2013,sarapata2014}. While such models are employed widely, they are ill-suited for describing extinction of finite populations, as the population is not restricted to integer states, that is, whole numbers of individuals within the population. \textit{Discrete} models enforce the number of individuals to be an integer, as each individual in the model is simulated explicitly \cite{codling2008}. Further, discrete models have the benefit that biological mechanisms, such as birth, death, and movement, can be directly imposed at the level of individuals \cite{baker2010,codling2008,johnston2017,simpson2011}. \\

A popular choice of discrete model for describing population dynamics is the lattice-based random walk \cite{baker2010,bubba2019,codling2008,el2019,fadai2019,field2010,johnston2017,simpson2011}. Individuals in the model reside on a pre-defined lattice, and undergo processes such as birth, death, and movement in a stochastic manner according to biologically-inspired rules \cite{baker2010,codling2008,fadai2019,johnston2017,simpson2011}. The lattice provides a natural method for incorporating population limits and finite size effects, known as an exclusion process, by imposing the restriction that each lattice site contains, at most, one individual \cite{chowdhury2005,embacher2018}. These lattice-based random walks have been successfully used to investigate processes in cell biology and ecology, amongst others \cite{bubba2019,fadai2019,johnston2014,johnston2014b}. \\

While the simplicity and flexibility of the lattice-based random walk is appealing, there are limitations to this modelling framework. In particular, the stochastic nature of the random walk means that a considerable number of simulations of the random walk process must be performed to obtain representative average behaviour of the population. This can be extremely computationally demanding, particularly if the random walk model is employed to perform parameter inference \cite{johnston2014b,johnston2016}. Furthermore, the random walk model is typically analytically intractable, and the amount of analytic insight is limited. To address this, continuum approximations of random walk processes have been developed \cite{baker2010,codling2008,simpson2011,johnston2016b}. Provided a judicious choice of approximation is made, typically depending on the rates of birth, death, and movement, these approximations accurately describe the average behaviour in the random walk \cite{baker2010,simpson2011,johnston2016b}. However, as noted previously, standard continuum models are ill-suited for describing population extinction. Therefore, the development of a continuum approximation that avoids the computational burden of repeated simulations of a stochastic random walk process, while faithfully capturing the behaviour in the random walk and accurately predicting population extinction would be beneficial. \\

Here we introduce a new PDE approximation of the underlying birth-death-movement lattice-based random walk, which we refer to as the state space diffusion approximation (SSDA). The SSDA involves two major components. First, the underlying random walk is approximated via a state space representation. The approximate state space of this random walk corresponds to the number of sites occupied by individuals. Second, the state space representation is approximated via the diffusion method, which is also known as the Fokker-Planck approximation \cite{doering2005,gardiner1985}. In contrast to other PDE approximations of lattice-based random walks, the dependent variables in the PDE are time and \textit{state space}, rather than time and physical space. As such, the solution to the PDE can be interpreted as the distribution of the number of occupied sites at a particular time.  Here we consider two mechanisms for birth and death \cite{fadai2019,johnston2017} : 
\begin{enumerate}[label=(\roman*)]
\item Logistic growth, where the intrinsic rates of birth and death of an individual are independent of the neighbouring lattice sites, and;
\item Allee growth, where the intrinsic rates of birth and death depend on whether an individual has zero occupied neighbouring sites, or at least one occupied neighbour site. 
\end{enumerate}
The logistic growth model is widely used to describe populations where low-density growth is close to exponential but the population eventually tends to a finite carrying capacity density due to competition for a shared resource \cite{murray2003}. Note that growth refers to the net birth/death rate of the population whereas intrinsic rates of birth/death refers to the birth/death of an individual. The logistic growth model has been used to describe the dynamics of populations of tumour cells \cite{gerlee2013,sarapata2014}, spruce budworms \cite{murray2003} and jellyfish polyps \cite{melica2014}, amongst others. The logistic model is relatively simple, but still captures key features of population behaviour. The Allee growth model is slightly more complicated, and describes population dynamics where growth is either inhibited or negative below a threshold population density \cite{courchamp1999,taylor2005,ovaskainen2001,johnson2019,neufeld2017}. This reduced growth rate can represent mechanisms where a threshold number of individuals are needed to maintain a population, such as for effective predator avoidance or for a viable number of breeding pairs \cite{courchamp1999,taylor2005}. Reduced or negative growth rates have been observed in populations of cancer cells \cite{johnson2019,neufeld2017}, as well as in animal populations, such as the gypsy moth \cite{johnson2006}. The widespread relevance of both the logistic and Allee growth models ensure that insight gleaned from analysis of population extinction in such models will be instructive in a range of research areas. \\

For both the logistic and Allee models, we demonstrate that the SSDA faithfully predicts the behaviour of the random walk process, provided that the standard mean-field assumption is satisfied. Further, as the SSDA explicitly accounts for the possibility that an individual random walk will undergo extinction, we demonstrate that the SSDA provides a more accurate approximation of the average random walk behaviour compared to standard mean-field ODEs. The SSDA provides an estimate of both the rate of extinction and the state space distribution, and we show that these estimates are consistent with the underlying random walk process. Finally, we highlight that the time taken to obtain the solution to the SSDA is orders of magnitude lower than the time taken to perform sufficiently many realisations of the random walk.

\section{Model}
\subsection{Logistic Model}

\begin{figure}
\begin{center}
\includegraphics[width=0.9\textwidth]{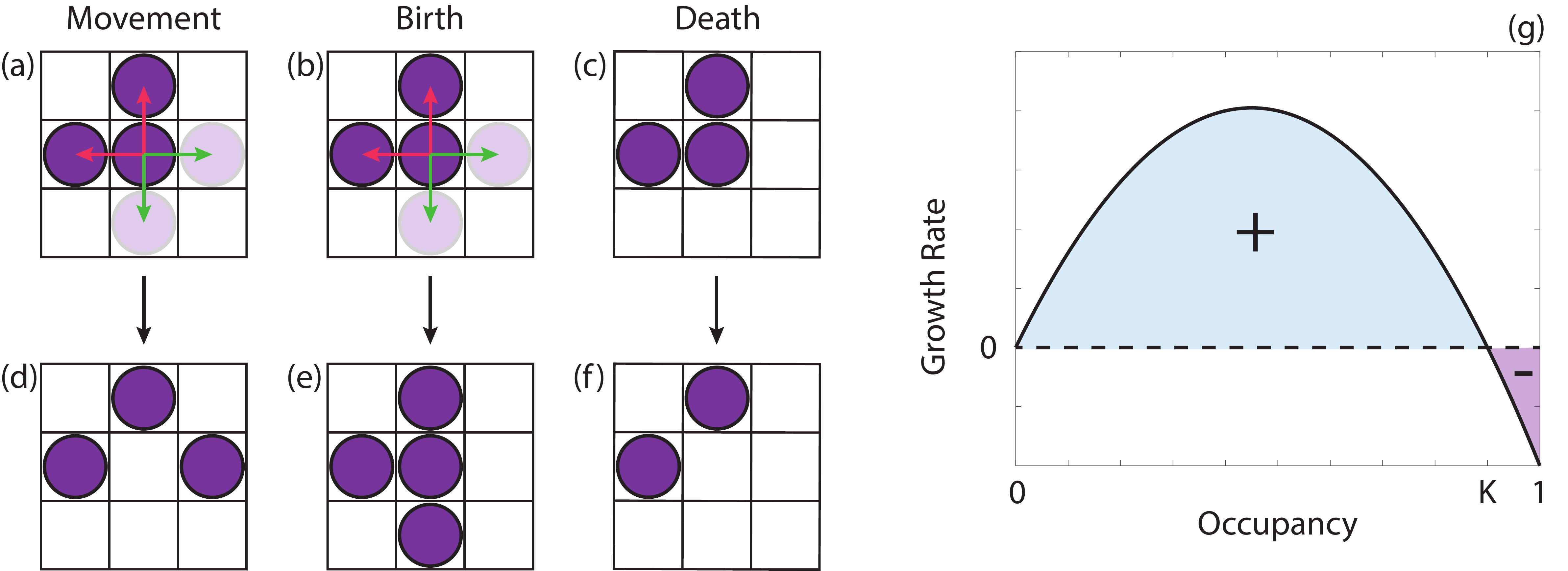}
\caption{Mechanisms in the logistic random walk model. (a) Potential movement events for the individual in the centre of the lattice. (b) Potential birth events for the individual in the centre of the lattice. Green arrows indicate directions that would result in a successful event, red arrows indicate directions that would result in an unsuccessful event. (c) Potential death events for the individual in the centre of the lattice. (d)-(f) Lattice configuration after the corresponding event in (a)-(c). (g) Typical growth rate curve as a function of the occupancy of the lattice. Cyan regions correspond to positive growth and purple regions correspond to negative growth. $K = (P_b-P_d)/P_b$ is the carrying capacity.}
\label{F0p5}
\end{center}
\end{figure}

Consider an $n$-dimensional lattice-based birth-death-movement random walk with exclusion on a lattice with $N$ sites \cite{baker2010,chowdhury2005}. While, in theory, $n$ is arbitrary as continuum approximations of the random walk under spatially-uniform initial conditions are independent of $n$, the most relevant $n$ for biological and ecological processes are $n = 1,2,3$. Here we present results corresponding to a square lattice with $n =2$. Individuals in the random walk undergo birth, death and movement events at rates per unit time $P_b$, $P_d$ and $P_m$, respectively \cite{baker2010,markham2013} (Figure \ref{F0p5}(a)-(f)). During a birth event, we select an individual at random, and select one of $2n$ nearest-neighbour sites in the von Neumann neighbourhood of the selected individual. The individual attempts to place a daughter individual at the target site and is successful provided that the target site is unoccupied; otherwise the event is aborted. Similarly, during a movement event, we select an individual at random, as well as a nearest-neighbour site. The individual attempts to move to the target site, and the event is only successful if the target site is unoccupied. During a death event, an individual is selected at random and removed from the lattice. Initially, $N_0$ sites are selected at random, and an individual is placed at each selected site. This type of random walk process has been examined previously \cite{baker2010,markham2013}. For spatially-uniform initial conditions, periodic boundary conditions, and under the mean-field assumption, where the probability that a particular lattice site is occupied is independent of all other sites, the evolution of the average proportion of occupied sites $0 \leq \overline{S} \leq 1$ is given by \cite{baker2010,markham2013}
\begin{equation}
\frac{\text{d}\overline{S}}{\text{d}t} = P_b\overline{S}(1-\overline{S}) - P_d\overline{S}, \label{eq:LogisticODE}
\end{equation}
as highlighted in Figure \ref{F0p5}(g). \\

Here we instead consider the probability, $f(S,t)$, that the system has $S$ occupied sites at time $t$. The system evolves in terms of the state variable $S$ during a single time step of duration $\tau$, denoted $\delta f(S,t) = f(S,t+\tau)-f(S,t)$, according to
\begin{align}
\delta f(S,t) &= P_b\left(\frac{S-1}{N}\right)\left(1 - \frac{S-2}{N-1}\right)f(S-1,t) +  P_d\left(\frac{S+1}{N}\right)f(S+1,t) \nonumber \\ & - P_b\frac{S}{N}\left(1 - \frac{S-1}{N-1}\right)f(S,t) - \frac{1}{N}P_dSf(S,t). \label{eq:LogisticTransition}
\end{align}
The first term corresponds to the transition from state $S-1$ to state $S$ due to an individual undergoing a birth event. This occurs proportional to the birth rate, $P_b$, the proportion of the lattice that is occupied by individuals, $(S-1)/N$, the probability that a nearest-neighbour site is unoccupied, $1-(S-2)/(N-1)$, and the probability that the system is in state $S-1$, \\ $f(S-1,t)$. We note that this term involves the standard mean-field assumption that the probability that a site is occupied by an individual is independent of whether a nearest-neighbour site is occupied \cite{baker2010}. The second term corresponds to the transition from state $S+1$ to state $S$ due to an individual undergoing a death event. This occurs proportional to the death rate, $P_d$, the proportion of the lattice that is occupied by individuals, $(S+1)/N$, and the probability that the system is in state $S+1$, $f(S+1,t)$. The third and fourth terms represent events that result in the system moving from state $S$ to states $S+1$ and $S-1$, corresponding to birth and death events, respectively. Note that the rate of movement, $P_m$, does not appear in Equation \eqref{eq:LogisticTransition}, as movement events do not change the number of individuals in the system. Instead, the ratio of $P_m/P_b$ and $P_m/P_d$ influences whether the assumption of independence between the occupancy of nearest-neighbour sites is appropriate \cite{baker2010}. \\

The system has two boundary cases that occur at $S = 1$, where
\begin{align*}
\delta f(1,t) &= \frac{2}{N}P_df(2,t) - \frac{1}{N}(P_b+P_d)f(1,t),
\end{align*}
implying that probability mass is lost to the system at $S = 1$ corresponding to $P_df(1,t)/N$; and $S = N$, where
\begin{align*}
\delta f(N,t) &= P_b\bigg(\frac{N-1}{N}\bigg)\bigg(1-\frac{N-2}{N-1}\bigg)f(N-1,t) - P_df(N,t),
\end{align*}
and hence no probability mass passes through $S = N$. \\

Introducing a change of variables $S = Ns$, where $1/N \leq s \leq 1$ is the proportion of occupied sites \cite{chalub2011,chalub2014}, the system becomes
\begin{align*}
\delta f(s,t) &= P_b\frac{N}{N-1}\left(s-\frac{1}{N}\right)\left(1 - s + \frac{1}{N}\right)f\left(s-\frac{1}{N},t\right) +  P_d\left(s+\frac{1}{N}\right)f\left(s+\frac{1}{N},t\right) \\ & - P_b\frac{N}{N-1}s\left(1 - s\right)f(s,t) - P_dsf(s,t).
\end{align*}

\noindent Expanding $f(s,t)$ in a Taylor series, neglecting terms $O(N^{-3})$, and dividing by the time step $\tau$, we obtain
\begin{align*}
\frac{\delta f(s,t)}{\tau} &= \frac{1}{2N^2\tau}\left[P_b\frac{N}{N-1}\left(s-\frac{1}{N}\right)\left(1 - s + \frac{1}{N}\right) +  P_d\left(s+\frac{1}{N}\right)\right]\frac{\partial^2 f(s,t)}{\partial s^2} \\ & - \frac{1}{N\tau}\left[P_b\frac{N}{N-1}\left(s-\frac{1}{N}\right)\left(1 - s + \frac{1}{N}\right) -  P_d\left(s+\frac{1}{N}\right)\right]\frac{\partial f(s,t)}{\partial s} \\ & + \frac{1}{\tau}\left[P_b\frac{N}{N-1}\left(2s-1-\frac{1}{N}\right) + P_d\right]f(s,t) + O(N^{-3}),
\end{align*}
or in conservative form,
\begin{align*}
\frac{\delta f(s,t)}{\tau} &= \frac{1}{N\tau}\frac{\partial}{\partial s}\left(\frac{1}{2N}\left[P_b\frac{N}{N-1}\left(s-\frac{1}{N}\right)\left(1 - s + \frac{1}{N}\right) +  P_d\left(s+\frac{1}{N}\right)\right]\frac{\partial f(s,t)}{\partial s} \right. \\ & \left. - \left[P_b\frac{N}{N-1}\left(s-\frac{1}{N}\right)\left(1 - s + \frac{1}{N}\right) -  P_d\left(s+\frac{1}{N}\right) \right. \right. \\ & \left. \left. + \frac{1}{2N}\left\{P_b\frac{N}{N-1}\left(1 - 2s + \frac{2}{N}\right) + P_d\right\}\right]f(s,t)\right).
\end{align*}
%P_b\frac{N}{N-1}\left(s-\frac{1}{N}\right)\left(1 - s + \frac{1}{N}\right)C\left(s-\frac{1}{N},t\right) +  P_d\left(s+\frac{1}{N}\right)C\left(s+\frac{1}{N},t\right) \\ & - P_b\frac{N}{N-1}s\left(1 - s\right)f(s,t) - P_dsf(s,t).
Taking the limit $N\to\infty$ and $\tau\to 0$ jointly such that $N\tau$ is constant \cite{chalub2009}, and noting that, in practice, $N$ and $\tau$ are finite, we obtain
\begin{align*}
\frac{\partial f(s,t)}{\partial t}&= \frac{\partial}{\partial s}\left(\frac{1}{2N}\left[P_b\frac{N}{N-1}\left(s-s^2+\frac{2s}{N}\right) +  P_d\left(s+\frac{1}{N}\right)\right]\frac{\partial f(s,t)}{\partial s} \right. \\ & \left. - \left[P_b\frac{N}{N-1}\left(s-\frac{1}{N}\right)\left(1 - s + \frac{1}{N}\right) -  P_d\left(s+\frac{1}{N}\right) \right. \right. \\ & \left. \left. + \frac{1}{2N}\left\{P_b\frac{N}{N-1}\left(1 - 2s + \frac{2}{N}\right) + P_d\right\}\right]f(s,t)\right).
\end{align*}
Simplifying, the PDE gives
\begin{align}
\frac{\partial f(s,t)}{\partial t}= \frac{\partial}{\partial s}\left(\frac{1}{2N}\bigg[a(s) +  b(s)\bigg]\frac{\partial f(s,t)}{\partial s} + \left[b(s) - a(s) - \frac{1}{2N}\left\{a'(s) + b'(s)\right\}\right]f(s,t)\right)\label{eq:LogisticPDE} , 
\end{align}
where 
\begin{equation}
a(s) = P_b\frac{N}{N-1}\left(s-\frac{1}{N}\right)\left(1 - s + \frac{1}{N}\right),
\end{equation}
and
\begin{equation}
b(s) = P_d\left(s+\frac{1}{N}\right),
\end{equation}
noting that terms $O(N^{-3})$ are neglected. The boundary conditions are given by
\begin{align*}
\frac{1}{2N}\bigg[a(s) +  b(s)\bigg]\frac{\partial f(s,t)}{\partial s} + \left[b(s) - a(s) - \frac{1}{2N}\left\{a'(s) + b'(s)\right\}\right]f(s,t) = \frac{P_d}{N}f(s,t) \ \text{at} \ s = 1/N, 
\end{align*}
and
\begin{align*}
\frac{1}{2N}\bigg[a(s) +  b(s)\bigg]\frac{\partial f(s,t)}{\partial s} + \left[b(s) - a(s) - \frac{1}{2N}\left\{a'(s) + b'(s)\right\}\right]f(s,t) = 0 \ \text{at} \ s = 1. 
\end{align*}

We note that the approach employed here is similar to obtaining diffusion approximations of infectious disease processes, where the size of the state space renders solution impractical \cite{chalub2011,chalub2014,doering2005,rebuli2017}. However, to the best of our knowledge, this type of approach has not been employed to analyse and approximate lattice-based random walks with exclusion. \\

Consistent with the hyperbolic scaling imposed in the SSDA, we expect advection to provide the dominant contribution to transport of the probability distribution. That is, we expect the initial condition to evolve primarily according to the advective velocity $b(s)-a(s)$. While this does neglect the contribution of diffusive transport, this approximation is appropriate provided that $N$ is large, as the diffusive contribution is inversely proportional to $N$ (see Equation \eqref{eq:LogisticPDE}). As such, we expect the mode of the probability distribution, $s_{\text{mo}}$, to evolve from the initial location of the peak of the initial condition to the stable steady state of 
\begin{equation*}
\frac{\text{d}s_{\text{mo}}}{\text{d}t} = a(s_{\text{mo}})-b(s_{\text{mo}}). 
\end{equation*}
Note that this is similar to the standard mean-field ODE, but that this describes the \textit{mode} of the probability distribution, rather than the \textit{mean} of the probability distribution.
% If we consider the limit where $N \to \infty$, the diffusion terms in Equation \eqref{eq:LogisticPDE} disappear, and we obtain
%\begin{equation}
%\frac{\partial f(s,t)}{\partial t} = \frac{\partial}{\partial s}\bigg(\left[b(s) - a(s)\right]f(s,t)\bigg). \label{eq:LogisticLimit}
%\end{equation}
%If the initial condition is a point source, then the mean value of $s$ evolves in time at a rate corresponding to the advective velocity $a(s)-b(s)$, and hence the mean value of $s$ obtained from Equation \eqref{eq:LogisticLimit} is equivalent to that obtained from the ODE describing the evolution of the average proportion of occupied sites, Equation \eqref{eq:LogisticODE}.
\subsection{Allee Model}

\begin{figure}
\begin{center}
\includegraphics[width=1.0\textwidth]{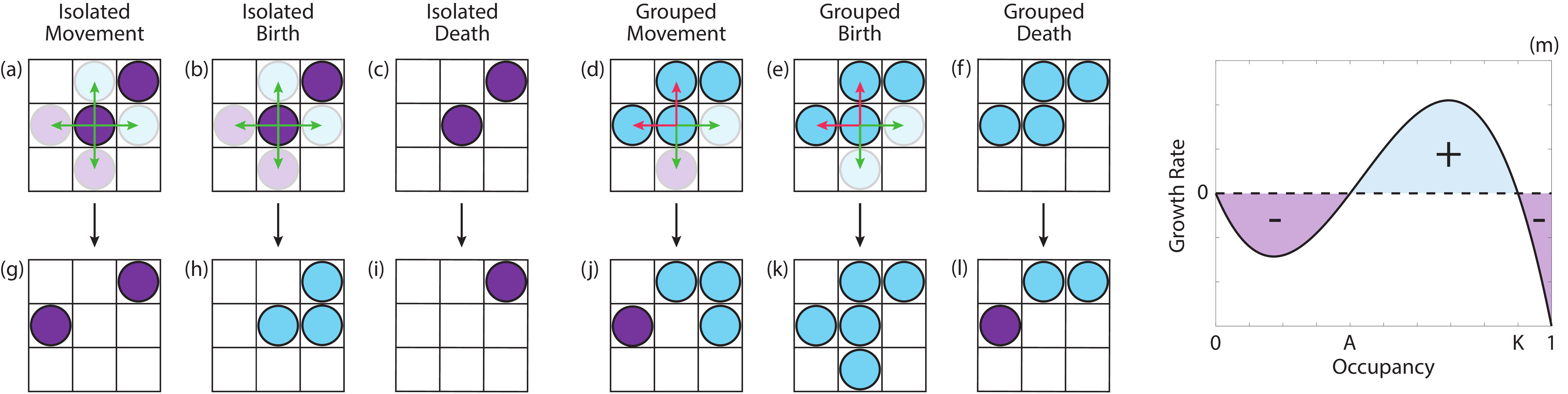}
\caption{Mechanisms in the Allee random walk model. (a)-(c) Potential movement, birth and death events, respectively, for the isolated individual in the centre of the lattice. (d)-(f) Potential movement, birth and death events, respectively, for the grouped individual in the centre of the lattice. Green arrows indicate directions that would result in a successful event, red arrows indicate directions that would result in an unsuccessful event. (g)-(l) Lattice configuration after the corresponding event in (a)-(f). Isolated individuals are shown in purple, grouped individuals are shown in cyan. (m) Typical growth rate curve as a function of the occupancy of the lattice. Cyan regions correspond to positive growth and purple regions correspond to negative growth. $A$ is the threshold parameter and $K$ is the carrying capacity.}
\label{F1}
\end{center}
\end{figure}

We now consider a two-dimensional birth-death-movement random walk with exclusion on a lattice with $N$ sites \cite{chowdhury2005}. Here the birth and death rates depend on whether an individual agent has zero nearest-neighbour agents or at least one nearest-neighbour agent, referred to as isolated and grouped individuals, respectively \cite{fadai2019,johnston2017} (Figure \ref{F1}(a)-(l)). These rates are $P_b^i$, $P_b^g$, $P_d^i$ and $P_d^g$ for birth and death of isolated and grouped individuals, respectively. This corresponds to either a competitive or co-operative process, depending on whether it is beneficial to be part of a group of individuals or to be isolated \cite{johnston2017}. It has been shown that in one dimension these birth and death mechanisms give rise to an Allee effect \cite{johnston2017} and, in two dimensions, these mechanisms give rise to a per capita growth rate that is qualitatively similar to the Allee effect \cite{fadai2019}. In certain parameter regimes, growth rates can be obtained that are qualitatively similar to the weak, strong and hyper Allee effects \cite{fadai2019}. The weak Allee effect refers to a per capita growth rate where growth is inhibited, but remains positive, at low individual densities \cite{johnson2019,taylor2005}. The strong Allee effect refers to the case where growth is negative below a threshold individual density, and positive otherwise \cite{dennis2002,taylor2005}. The hyper Allee effect refers to per capita growth rates that may have additional steady states \cite{fadai2019}. As the strong Allee effect is perhaps the more interesting model with respect to population extinction, as fluctuations in the number of individuals that take the population below the threshold density can result in extinction \cite{dennis2002}, we focus on parameter suites that correspond to the strong Allee effect, noting that it is straightforward to analyse other parameter regimes. In Figure \ref{F1}(m), we present the growth curve as a function of the proportion of occupied lattice sites. Under the standard mean-field assumption for a spatially-uniform initial condition and periodic boundary conditions, the evolution of the average proportion of occupied sites, $0 \leq \overline{S} \leq 1$, for the strong Allee effect is \cite{taylor2005,johnston2017}
\begin{equation}
\frac{\text{d}\overline{S}}{\text{d}t} = r\overline{S}(K-\overline{S})(C(\overline{S})-C(A)), \label{eq:AlleeODE}
\end{equation}
where $r > 0$ is the net birth rate, $0 \leq K \leq 1$ is the carrying capacity, $0 < A < K$ is the threshold parameter, and $C(\overline{S})$ satisfies $C(\overline{S}) < C(A)$ for $0 \leq \overline{S} < A$, and $C(\overline{S}) > C(A)$ otherwise. \\

For the strong Allee effect, it is known that the standard mean-field ODE is inaccurate near the unstable steady state, where the growth rate transitions from negative to positive \cite{fadai2019}. As above, we consider how the system evolves in terms of the state variable, the number of occupied sites, during a single time step:
\begin{align*}
\delta f(S,t) &= P_b^g\frac{S-1}{N}\left(1-\frac{S-2}{N-1}\right)\left(1-\left[1-\frac{S-2}{N-2}\right]\left[1-\frac{S-2}{N-3}\right]\left[1-\frac{S-2}{N-4}\right]\right)f(S-1,t) \\ & - P_b^g\frac{S}{N}\left(1-\frac{S-1}{N-1}\right)\left(1-\left[1-\frac{S-1}{N-2}\right]\left[1-\frac{S-1}{N-3}\right]\left[1-\frac{S-1}{N-4}\right]\right)f(S,t) \\ & + P_b^i\frac{S-1}{N}\left(1 - \frac{S-2}{N-1}\right)\left(1 - \frac{S-2}{N-2}\right)\left(1 - \frac{S-2}{N-3}\right)\left(1 - \frac{S-2}{N-4}\right)f(S-1,t) \\ & - P_b^i\frac{S}{N}\left(1 - \frac{S-1}{N-1}\right)\left(1 - \frac{S-1}{N-2}\right)\left(1 - \frac{S-1}{N-3}\right)\left(1 - \frac{S-1}{N-4}\right)f(S,t) \\ & +P_d^g \frac{S+1}{N}\left(1 - \left[1 - \frac{S}{N-1}\right]\left[1 - \frac{S}{N-2}\right]\left[1 - \frac{S}{N-3}\right]\left[1 - \frac{S}{N-4}\right]\right)f(S+1,t) \\ & - P_d^g \frac{S}{N}\left(1 - \left[1 - \frac{S-1}{N-1}\right]\left[1 - \frac{S-1}{N-2}\right]\left[1 - \frac{S-1}{N-3}\right]\left[1 - \frac{S-1}{N-4}\right]\right)f(S,t) \\ & + P_d^i\frac{S+1}{N}\left(1-\frac{S}{N-1}\right)\left(1-\frac{S}{N-2}\right)\left(1-\frac{S}{N-3}\right)\left(1-\frac{S}{N-4}\right)f(S+1,t) \\ & - P_d^i\frac{S}{N}\left(1-\frac{S-1}{N-1}\right)\left(1-\frac{S-1}{N-2}\right)\left(1-\frac{S-1}{N-3}\right)\left(1-\frac{S-1}{N-4}\right)f(S,t).
\end{align*}
Note that if $P_b^g = P_b^i$ and $P_d^g = P_d^i$, the above system reduces to the logistic model. Following the same limit approach as previously, we obtain the following PDE describing the distribution of the state of the random walk process
\begin{align}
\frac{\partial f(s,t)}{\partial t}&= \frac{\partial}{\partial s}\left(\frac{1}{2N}\bigg[a(s) +  b(s)\bigg]\frac{\partial f(s,t)}{\partial s} + \left[b(s) - a(s) - \frac{1}{2N}\left\{a'(s) + b'(s)\right\}\right]f(s,t)\right), \label{eq:AlleePDE}
\end{align}
where 
\begin{align}
a(s) &= P_b^g\frac{N^4}{(N-1)(N-2)(N-3)(N-4)}\left(s-\frac{1}{N}\right)\left(1 - s + \frac{1}{N}\right) \nonumber \\ & \times \left(\frac{(N-2)(N-3)(N-4)}{N^3}-\bigg[1-s\bigg]\left[1-s-\frac{1}{N}\right]\left[1-s-\frac{2}{N}\right]\right) \nonumber \\ & + P_b^i\frac{N^4}{(N-1)(N-2)(N-3)(N-4)} \nonumber \\ & \times \left(s-\frac{1}{N}\right)\left(1 - s + \frac{1}{N}\right)\bigg(1-s\bigg)\left(1-s-\frac{1}{N}\right)\left(1-s-\frac{2}{N}\right),
\end{align}
and
\begin{align}
b(s) &=P_d^g\frac{N^4}{(N-1)(N-2)(N-3)(N-4)}\left(s+\frac{1}{N}\right) \nonumber \\ & \times \left(\frac{(N-1)(N-2)(N-3)(N-4)}{N^4}-\left[1-s-\frac{1}{N}\right]\left[1-s-\frac{2}{N}\right]\left[1-s-\frac{3}{N}\right]\left[1-s-\frac{4}{N}\right]\right) \nonumber \\ & + P_d^i\frac{N^4}{(N-1)(N-2)(N-3)(N-4)} \nonumber \\ & \times \left(s+\frac{1}{N}\right)\left(1 - s - \frac{1}{N}\right)\left(1-s-\frac{2}{N}\right)\left(1-s-\frac{3}{N}\right)\left(1-s-\frac{4}{N}\right).
\end{align}
The boundary conditions are given by
\begin{align*}
\frac{1}{2N}\bigg[a(s) +  b(s)\bigg]\frac{\partial f(s,t)}{\partial s} + \left[b(s) - a(s) - \frac{1}{2N}\left\{a'(s) + b'(s)\right\}\right]f(s,t) = P_d^if(s,t) \ \text{at} \ s = 1/N,
\end{align*}
and
\begin{align*}
\frac{1}{2N}\bigg[a(s) +  b(s)\bigg]\frac{\partial f(s,t)}{\partial s} + \left[b(s) - a(s) - \frac{1}{2N}\left\{a'(s) + b'(s)\right\}\right]f(s,t) = 0 \ \text{at} \ s = 1.
\end{align*}
Note that the boundary condition at $s = 1/N$ only depends on $P_d^i$ as it is impossible to have any grouped agents at this $s$ value, which corresponds to a single occupied lattice site. Similar to the logistic model, the standard approach to examine how the system evolves is to use a mean-field approximation, which results in the following ODE \cite{fadai2019,johnston2017}
\begin{equation*}
\frac{\text{d}\overline{S}}{\text{d}t} = P_b^g \overline{S}(1-\overline{S})(1-(1-\overline{S})^3) + P_b^i\overline{S}(1-\overline{S})^4 - P_d^g\overline{S}(1-(1-\overline{S})^4) - P_d^i\overline{S}(1-\overline{S})^4,
\end{equation*}
which is equivalent to Equation \eqref{eq:AlleeODE} in parameter regimes that correspond to the strong Allee effect. \\

In general, the form of the PDE arising from a lattice-based birth-death-movement random walk process is
\begin{align*}
\frac{\partial f(s,t)}{\partial t}&= \frac{\partial}{\partial s}\left(\frac{1}{2N}\bigg[a(s) +  b(s)\bigg]\frac{\partial f(s,t)}{\partial s} + \left[b(s) - a(s) - \frac{1}{2N}\left\{a'(s) + b'(s)\right\}\right]f(s,t)\right),
\end{align*}
where $a(s)$ are the coefficients of the $f(S-1,t)$ terms and $b(s)$ are the coefficients of the $f(S+1,t)$ terms, cast with respect of the scaled variable $s$ rather than $S$. That is, the coefficients represent the probability of transitioning from $S-1$ to $S$ and from $S+1$ to $S$, respectively.

\section{Results}

\begin{figure}
\begin{center}
\includegraphics[width=1.0\textwidth]{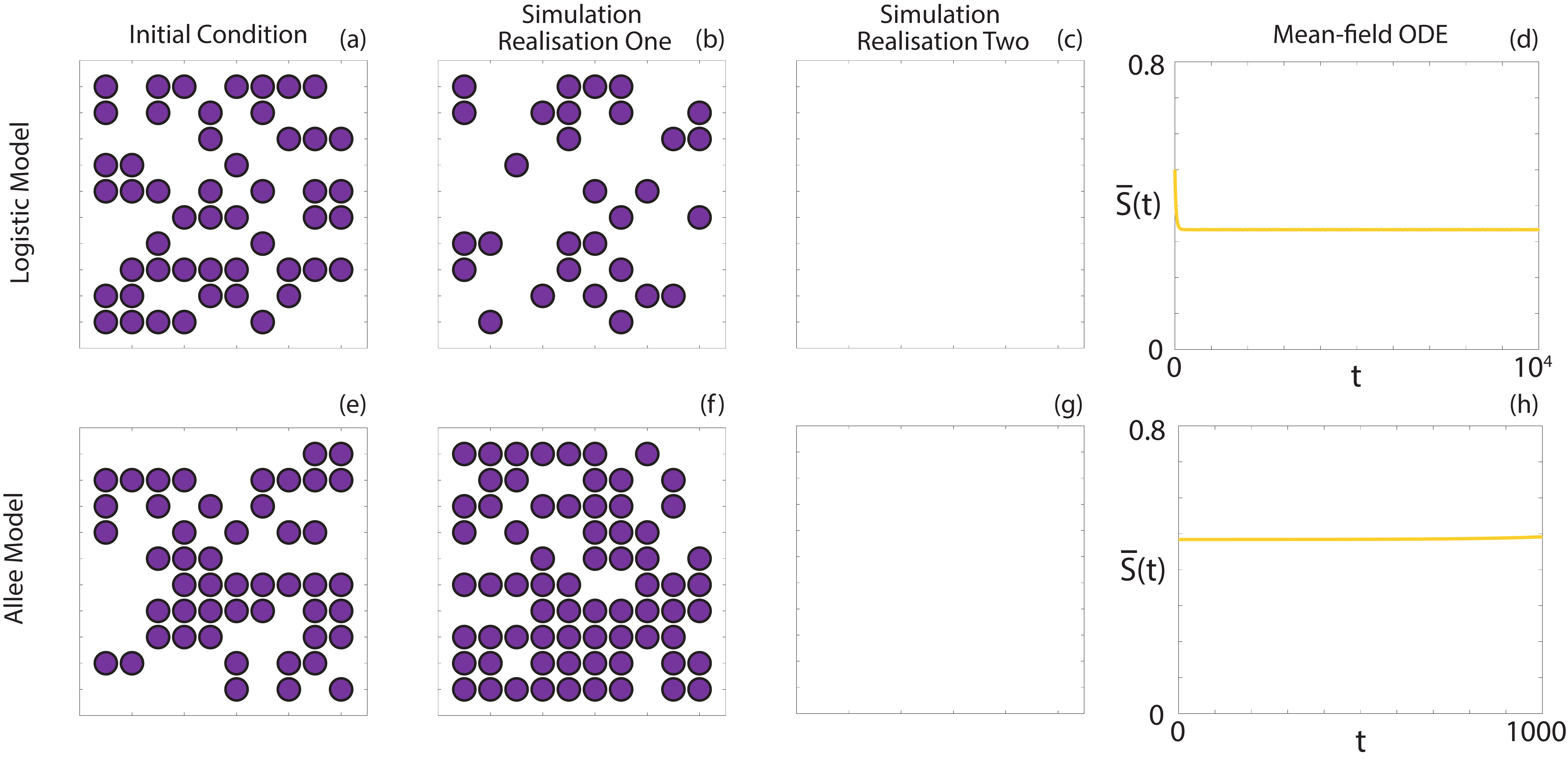}
\caption{Differences in the random walk obtained from identically-prepared initial conditions for (a)-(d) logistic growth model and the (e)-(h) Allee model. (a),(d) The initial condition in the random walk model. (b)-(c),(f)-(g) The final result of the random walk model for two realisations. (d),(h) The evolution of average occupancy obtained from the mean-field ODE. Parameters used are (a)-(d) $P_m = 1$, $P_b = 0.075$, $P_d = 0.05$, $T_{\text{end}} = 10^4$, $N = 100$, $N_0 = 50$ and (e)-(h) $P_m = 1$, $P_b^i = 0.01$, $P_b^g = 0.01$, $P_d^i = 0.04$, $P_d^g = 0.0025$, $T_{\text{end}} = 1000$, $N = 100$, $N_0 = 49$.}
\label{F1p5}
\end{center}
\end{figure}

We first highlight the stochastic nature of the random walk, and the resulting need for an approximation technique that provides the distribution of the number of occupied sites. In Figure \ref{F1p5}, we present simulation results for both logistic and Allee models where, for two identically-prepared simulations, two completely different lattice configurations arise at $t = T_{\text{end}}$. In one realisation, the population persists (Figures \ref{F1p5}(b),(f)), while in another realisation, the population undergoes extinction (Figures \ref{F1p5}(c),(g)). This information is not captured by the corresponding mean-field ODE, presented in Figures \ref{F1p5}(d),(h), which only provides information regarding the average occupancy. However, as detailed in Section 2, the SSDA explicitly accounts for population extinction and describes the distribution of the proportion of occupied sites. We refer to the latter throughout as the occupancy distribution. \\

\begin{figure}
\begin{center}
\includegraphics[width=1.0\textwidth]{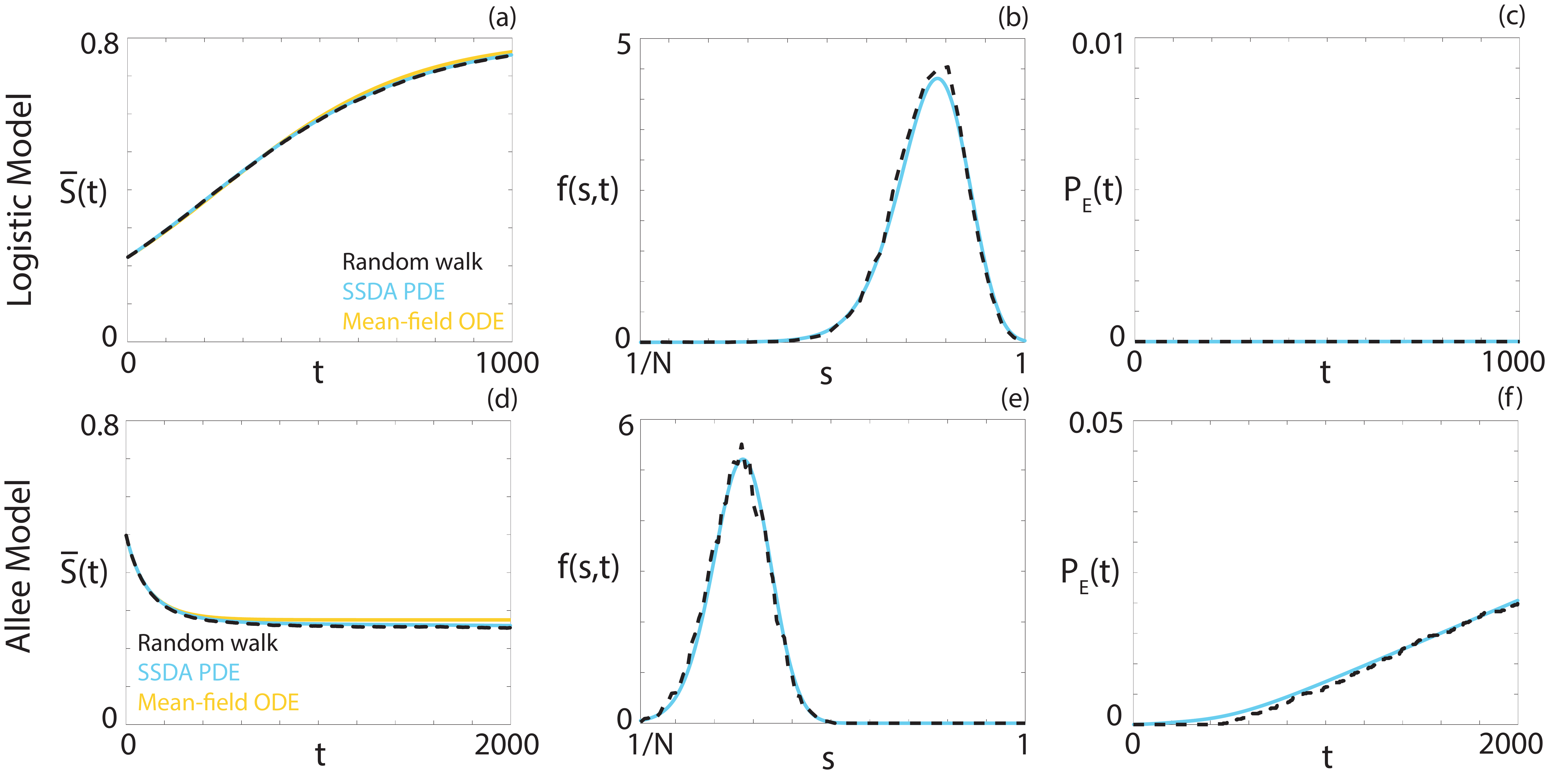}
\caption{Comparison between the average behaviour of the lattice-based random walk (black, dashed), the SSDA PDE (cyan) and the mean-field ODE (orange) for the (a)-(c) logistic growth model and the (d)-(f) Allee model. (a),(d) The evolution of the average occupancy. (b),(e) The occupancy distribution probability density function. (c),(f) The evolution of the extinction probability. Parameters used are (a)-(c) $P_m = 1$, $P_b = 0.005$, $P_d = 0.001$, $T_{\text{end}} = 1000$, $N = 36$, $N_0 =8$ and (d)-(f) $P_m = 1$, $P_b^i = 0.03$, $P_b^g = 0.01$, $P_d^i = 0.02$, $P_d^g = 0.01$, $T_{\text{end}} = 2000$, $N = 100$, $N_0 = 50$.}
\label{F2}
\end{center}
\end{figure}

We next verify that the predictions obtained from the SSDA are consistent with the average random walk behaviour and the mean-field ODE for parameter regimes where it is known that the mean-field approximation is appropriate. Specifically, these parameter regimes correspond to regimes where $P_m/P_b \gg 1$ and $P_m/P_d \gg 1$ \cite{baker2010}. Details of the numerical techniques employed to obtain solutions to the SSDA PDE are provided in Appendix A.  In Figure \ref{F2}(a),(d) we demonstrate that the evolution of the average occupancy obtained from the solution to the SSDA matches the average occupancy obtained from both the random walk and the mean-field ODE for both the logistic and Allee models. Results are obtained from $10^4$ identically-prepared realisations of the random walk. The average occupancy obtained from the SSDA is calculated via
\begin{equation*}
\overline{S}(t) = \sum_i s_i f(s_i,t) \Delta s,
\end{equation*}
where $s_i$ are the values of $s$ where a grid point is defined in the numerical solution, and $\Delta s$ is the space between grid points (Appendix A).
The SSDA provides additional information, compared to the mean-field ODE, with respect to the occupancy distribution, presented in Figure \ref{F2}(b),(e), and the probability that the population has undergone extinction by time $t$, $P_E(t)$, presented in Figure \ref{F2}(c),(f). The extinction probability is calculated via
\begin{equation*}
P_E(t) = \sum_i f(s_i,0) \Delta s - \sum_i f(s_i,t) \Delta s, 
\end{equation*}
that is, the difference in probability mass between the initial occupancy distribution and the occupancy distribution at time $t$. For both the occupancy distribution and the probability of population extinction, the predictions obtained from the SSDA match the average random walk behaviour. \\

\begin{figure}
\begin{center}
\includegraphics[width=1.0\textwidth]{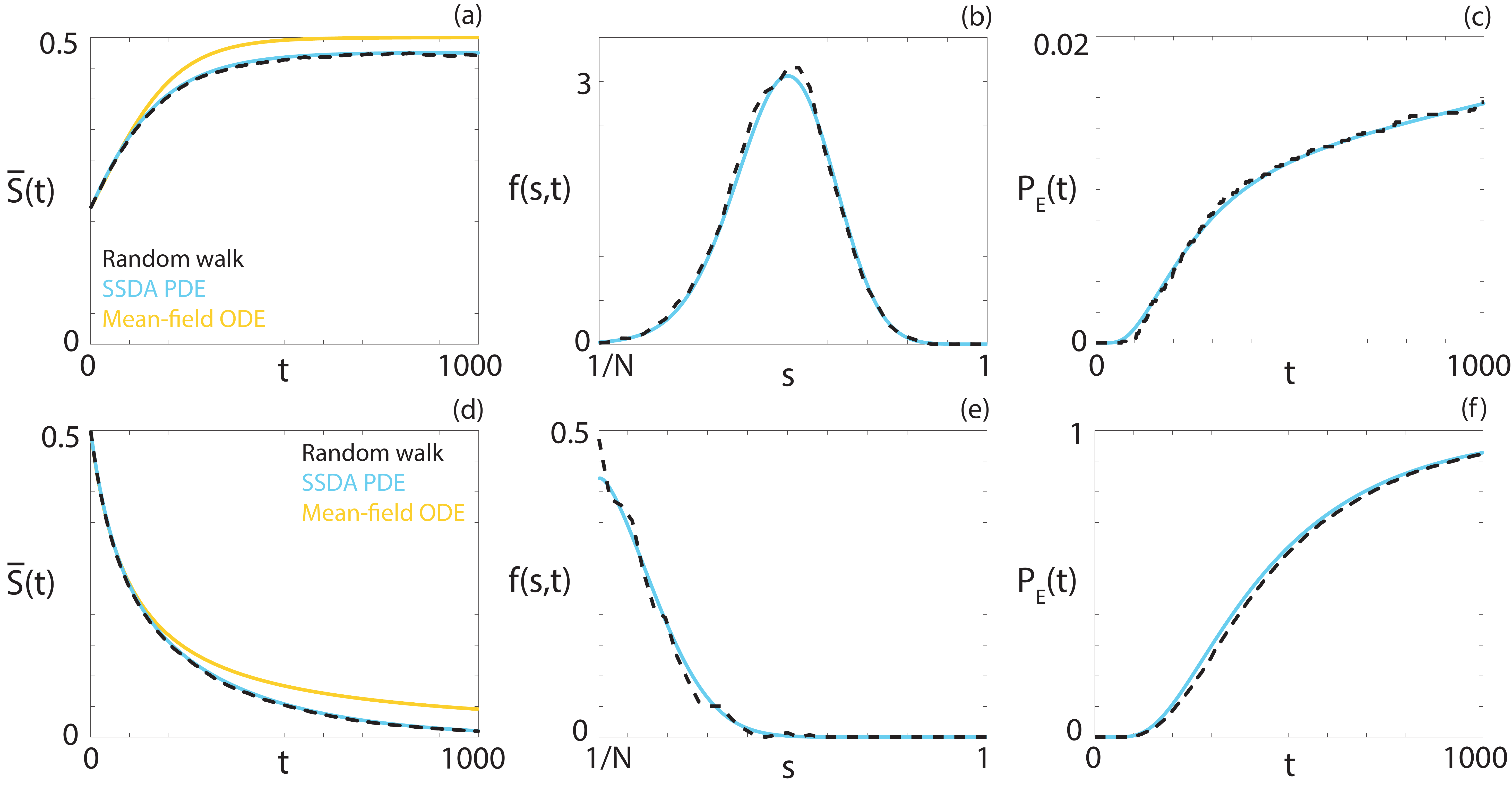}
\caption{Comparison between the average behaviour of the lattice-based random walk (black, dashed), the SSDA PDE (cyan) and the mean-field ODE (orange) for the logistic growth model. (a),(d) The evolution of the average occupancy. (b),(e) The occupancy distribution probability density function. (c),(f) The evolution of the extinction probability. Parameters used are $P_m = 1$, $P_b = 0.02$, $P_d = 0.01$, $T_{\text{end}} = 1000$, $N = 36$, $N_0 =8$ and (d)-(f) $P_m = 1$, $P_b = 0.02$, $P_d = 0.02$, $T_{\text{end}} = 1000$, $N = 36$, $N_0 = 18$.}
\label{F3}
\end{center}
\end{figure}

We now consider two parameter regimes for the logistic model where the possible extinction of the population has a significant impact, calculate the evolution of the average occupancy, the occupancy distribution and the probability of extinction, and present the results in Figure \ref{F3}. In all cases, results obtained from the SSDA match the average behaviour of the random walk well. In the first regime, presented in Figures \ref{F3}(a)-(c), the birth rate is twice the death rate, and hence the positive steady state in Equation \eqref{eq:LogisticODE} occurs at $\overline{S}  = 1/2$. Interestingly, despite the steady state existing well away from $\overline{S} = 0$, there is a non-zero chance that the population undergoes extinction. As such, the mean-field ODE overestimates the average occupancy in the random walk as time increases, as this approximation does not account for population extinction. At early time, it is unlikely that the population has undergone extinction, so the mean-field ODE provides a reasonable approximation. In contrast, the SSDA accounts for population extinction, and hence accurately reflects the average occupancy for all $t$ considered. The non-zero extinction chance can be observed in the occupancy distribution (Figure \ref{F3}(b)). The distribution is centred around $s = 1/2$ but the distribution is broad relative to the range of possible average occupancy values. As such, the flux through $s = 1/N$ is non-negligible. In the second parameter regime, presented in Figures \ref{F3}(d)-(f), there is no positive steady state solution to Equation \eqref{eq:LogisticODE}, and we expect the population to tend toward extinction. Both the mean-field ODE and the SSDA predict that the average occupancy will tend to zero; however, the mean-field ODE significantly overestimates the average occupancy compared to the SSDA. For example, at $t = 1000$, the mean-field ODE predicts an average occupancy that is approximately five times larger than that obtained from both the random walk and the SSDA. Further, it is unclear how the average occupancy is related to the probability that the population has undergone extinction, and hence the additional information provided by the SSDA is instructive in this regard. By $t = 1000$, there is approximately a 92\% chance that an individual population will have gone extinct and, for the non-extinct populations, the probability density of the occupancy distribution is monotonically decreasing with respect to the number of occupied sites. This indicates that for non-extinct populations, it is likely that the population is nearly extinct. This is in contrast to the results in Figure \ref{F3}(b), where non-extinct populations are not close to extinction. \\

\begin{figure}
\begin{center}
\includegraphics[width=1.0\textwidth]{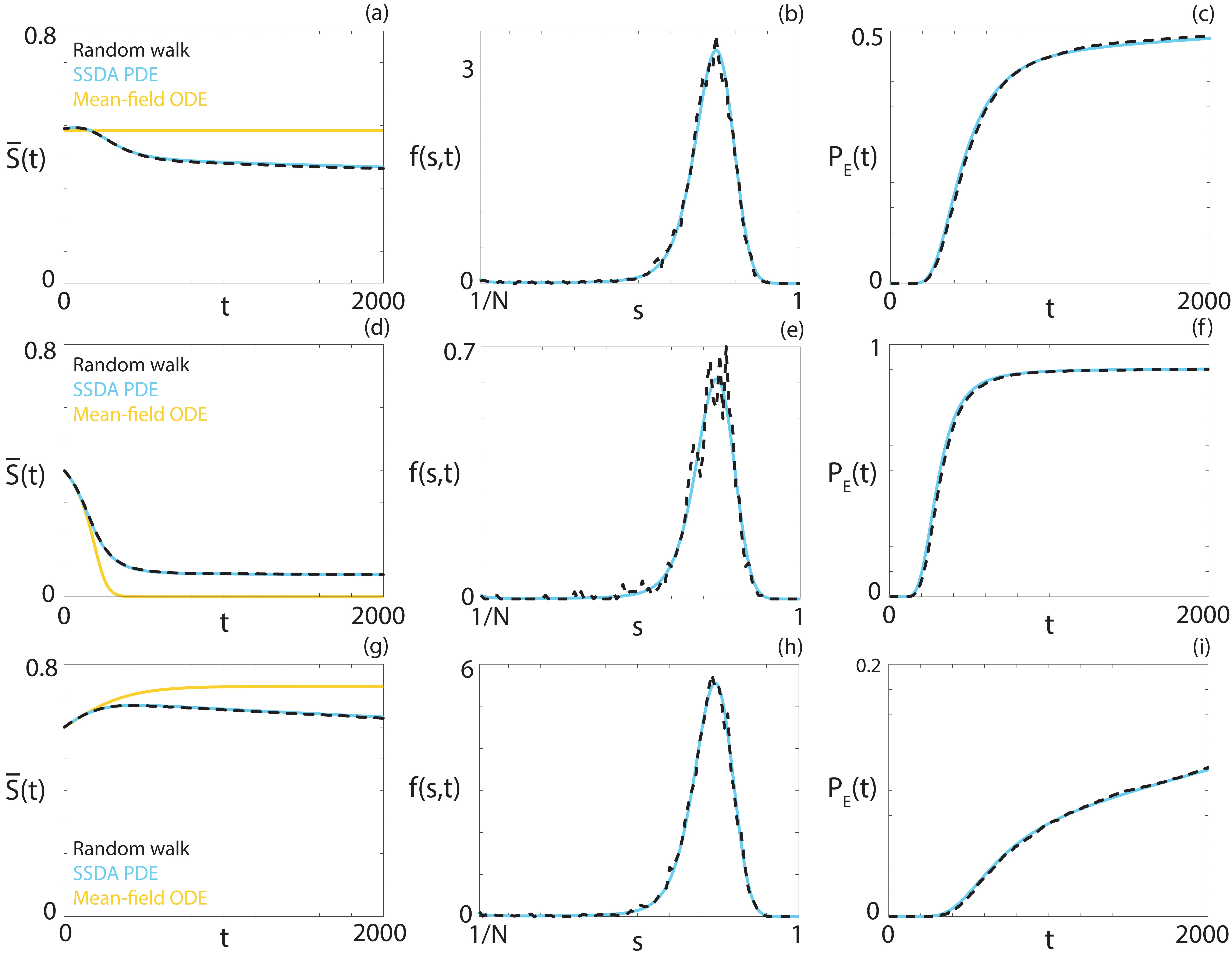}
\caption{Comparison between the average behaviour of the lattice-based random walk (black, dashed), the SSDA PDE (cyan) and the mean-field ODE (orange) for the Allee model. (a),(d),(g) The evolution of the average occupancy. (b),(e),(h) The occupancy distribution probability density function. (c),(f),(i) The evolution of the extinction probability. Parameters used are $P_m = 1$, $P_b^i = 0.01$, $P_b^g = 0.01$, $P_d^i = 0.04$, $P_d^g = 0.0025$, $T_{\text{end}} = 2000$, $N = 100$ and (a)-(c) $N_0 = 49$, (d)-(f) $N_0 = 40$, (g)-(i) $N_0 = 60$.}
\label{F4}
\end{center}
\end{figure}

Next, we consider three cases of the Allee model. In each case, we maintain the same rates of birth and death; only the number of initially occupied sites changes. The three different numbers of initially occupied sites correspond to initial average occupancies that are at ($\overline{S}(0) = A$, Figures \ref{F4}(a)-(c)), below ($\overline{S}(0) < A$, Figures \ref{F4}(d)-(f)) and above ($\overline{S}(0) > A$, Figures \ref{F4}(g)-(i)) the unstable steady state. As such, the mean-field ODE, Equation \eqref{eq:AlleeODE}, predicts that the evolution of the average occupancy should remain on the unstable steady state, tend towards extinction and tend towards the carrying capacity, respectively. \\

For the initial occupancy corresponding to $\overline{S}(0) = A$, we observe that the average occupancy obtained from both the SSDA and the random walk tends below $\overline{S} = A$ as time increases, but does not rapidly undergo extinction. This result is perhaps somewhat unintuitive based solely on the evolution of the average occupancy, but is clearer when considered in context of both the extinction probability and the occupancy distribution. Interestingly, the occupancy distribution is still centred around the carrying capacity $\overline{S} = K$, despite the average occupancy being below the unstable steady state. Combined with the probability of extinction at $t = 2000$ of approximately 50\%, this implies that in an individual realisation of a random walk one of two possibilities occur: either the population will become extinct or the population will tend to the carrying capacity. Due to this dichotomy, reporting the average occupancy is not an informative measure of average population behaviour. Instead, both the occupancy distribution and the extinction probability are required to understand the population behaviour. \\

Similar results are observed for the initial average occupancies corresponding to $\overline{S}(0) < A$ and $\overline{S}(0) > A$. The average occupancy in the case where $\overline{S}(0) < A$ appears to tend to a finite positive value in the random walk, and the corresponding SSDA. In contrast, the mean-field ODE predicts that the average occupancy will be near-zero by $t = 400$. Again, this discrepancy can be understood by considering the occupancy distribution, and the extinction probability. Here, a single population will become extinct approximately 90\% of the time, but will otherwise tend to the carrying capacity. This is again observed for the case where $\overline{S}(0) > A$, albeit with a much lower probability of extinction due to the higher initial occupancy. We observe that the occupancy distribution is extremely similar for all three initial occupancies, to a scaling factor. This suggests that provided a population does not go extinct, we expect the population to exist near the carrying capacity. Further, the probability that a population goes extinct depends on the initial occupancy. However, this dependence is less strict than in the standard Allee model, Equation \eqref{eq:AlleeODE}, where the population will tend to extinction with certainty if $\overline{S}(0) < A$ and will tend to the carrying capacity with certainty if $\overline{S}(0) > A$. \\

\begin{figure}
\begin{center}
\includegraphics[width=1.0\textwidth]{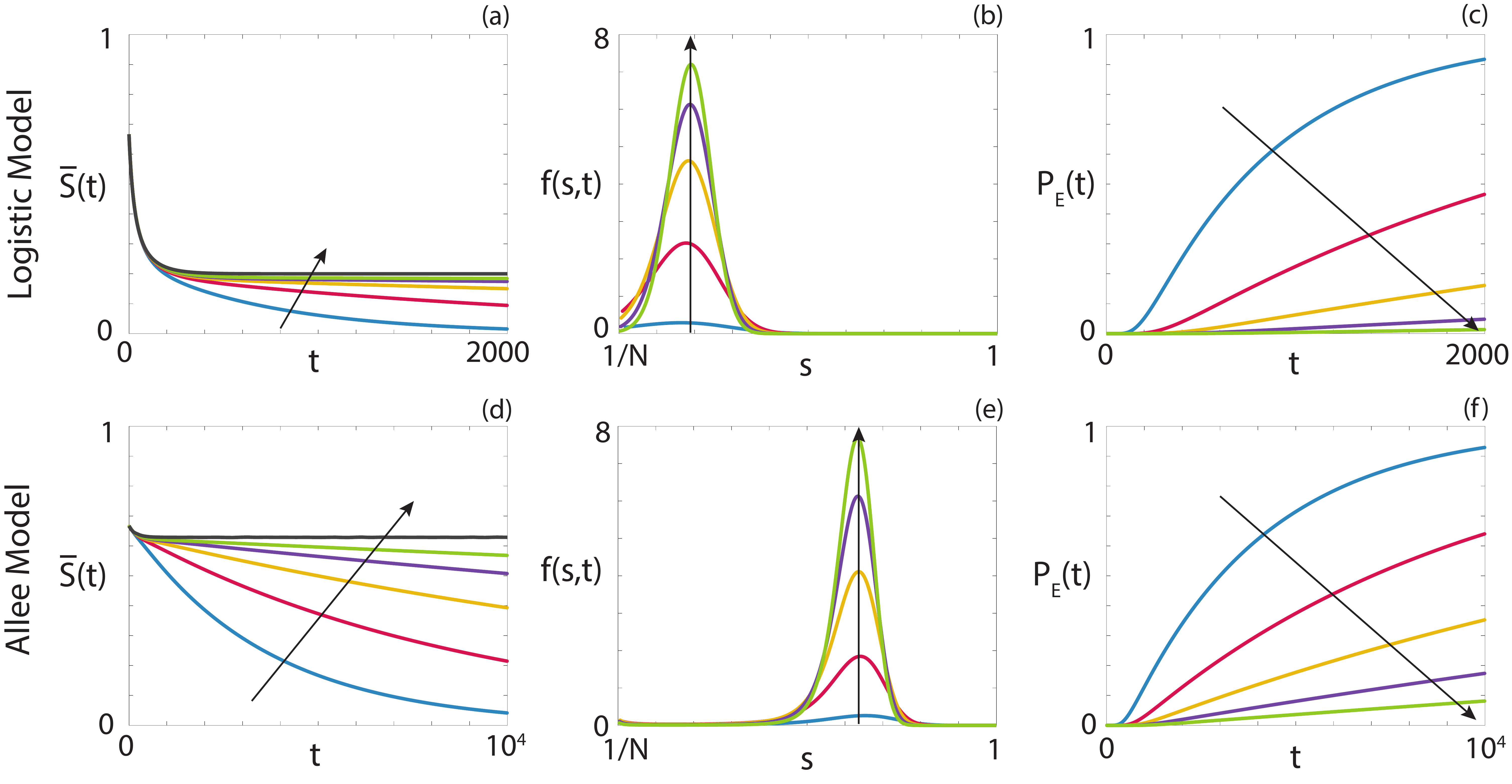}
\caption{SSDA predictions for the (a),(d) the evolution of the average occupancy, (b),(e) the occupancy distribution probability density function, and (c),(f) the evolution of the extinction probability for (a)-(c) the logistic model and (d)-(f) the Allee model for a range of $N$ values. Arrow indicates direction of increasing $N$. Parameters used are (a)-(c) $P_m = 1$, $P_b = 0.05$, $P_d = 0.04$, $T_{\text{end}} = 2000$ and (d)-(f) $P_m = 1$, $P_b^i = 0.02$, $P_b^g = 0.015$, $P_d^i = 0.04$, $P_d^g = 0.005$, $T_{\text{end}} = 10^4$. Coloured lines indicate different population sizes: (blue) $N = 60$, $N_0 = 40$, (red) $N = 120$, $N_0 = 80$, (orange) $N = 180$, $N_0 = 120$, (purple) $N = 240$, $N_0 = 180$, (green) $N = 300$, $N_0 = 240$. The grey line represents the mean-field ODE solution, that is, $N \to \infty$.}
\label{F5}
\end{center}
\end{figure}

As the difference between the SSDA and the standard mean-field approximation arises from whether the population size is finite or infinite, it is instructive to examine how different population sizes impact the average occupancy, occupancy distribution and the probability of extinction. We consider both the logistic model and the Allee model for a range of population sizes, with a consistent proportion of the lattice initially occupied. The parameter regimes considered have a stable steady state that is below the initial average occupancy but occurs at positive average occupancy. As such, we expect that the average occupancy should tend toward this positive value, aside from extinction events associated with finite population size. For the results obtained from the logistic model, presented in Figures \ref{F5}(a)-(c), we observe that the size of the population has a dramatic effect. For the smallest population $(N= 60)$, we observe that the average occupancy is close to zero by $t = 2000$ and that the extinction probability is approaching one. Increasing the population size, we observe that the average occupancy increases, corresponding to a decrease in extinction probability. Further, we see that the spread in the occupancy distribution decreases with an increase in population size. This corresponds to a smaller proportion of this distribution occurring near the boundary corresponding to extinction. The population size has a similar effect in the Allee model, as observed in the results presented in Figures \ref{F5}(d)-(f). As noted in the results presented in Figure \ref{F4}, provided that the population does not undergo extinction, the occupancy distribution is centred around the stable steady state. \\

\begin{figure}
\begin{center}
\includegraphics[width=1.0\textwidth]{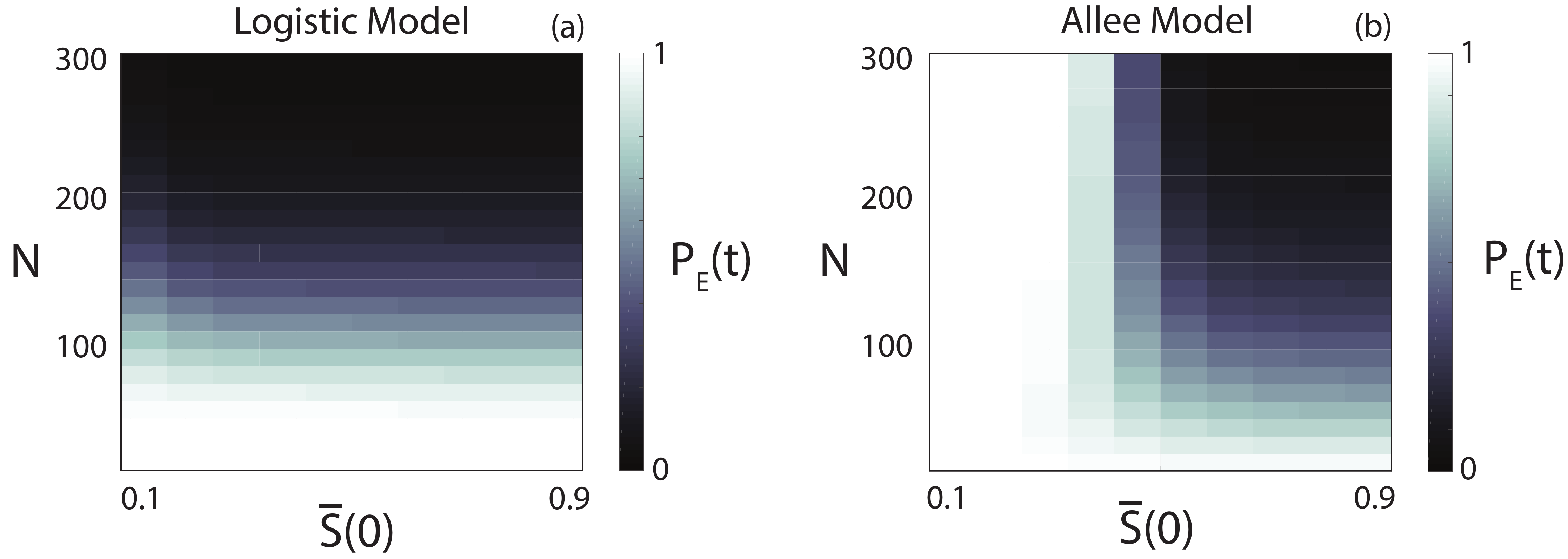}
\caption{SSDA predictions for the probability of population extinction by (a) $T_{\text{end}} = 2000$ and (b) $T_{\text{end}} = 10^4$ for (a) the logistic model and (b) the Allee model. Parameters used are (a) $P_m = 1$, $P_b = 0.05$, $P_d = 0.04$ and (b) $P_m = 1$, $P_b^i = 0.02$, $P_b^g = 0.015$, $P_d^i = 0.04$, $P_d^g = 0.005$.}
\label{F6}
\end{center}
\end{figure}

Finally, we evaluate the probability that a population will undergo extinction by a particular time for a suite of population sizes and initial average occupancies for both the logistic and Allee models, and present the results in Figure \ref{F6}. For the logistic model, we observe that increasing the population size corresponds to a decrease in the extinction probability. Interestingly, the initial occupancy is less important than the number of sites. This is consistent with the observation that the occupancy distribution will approach a quasi-steady state distribution on a faster timescale than extinction \cite{ovaskainen2001}. For the Allee model, the initial average occupancy is important due to the presence of the Allee threshold. If the initial average occupancy is below this threshold density, the population will tend to extinction. However, if the population size is sufficiently small, fluctuations in average occupancy can result in the population size crossing this threshold, and hence a portion of populations will tend to the carrying capacity. Similarly, if the initial average occupancy is above the Allee threshold density, the majority of populations will tend to the carrying capacity. The exact size of this majority depends on the distance between the initial average occupancy and the Allee threshold density, as well as the population size. This can be observed in Figure \ref{F6}(b), where increasing both the initial average occupancy and the population size results in a decrease in the probability of extinction. \\

\begin{table}
\begin{center}
\renewcommand{\arraystretch}{1.3}
\resizebox{1.0\textwidth}{!}{
\begin{tabular}{| l | c | c | c | c | c | c | c |}
\hline
Result & Fig. \ref{F2}(a)-(c) & Fig. \ref{F2}(d)-(f) & Fig. \ref{F3}(a)-(c) & Fig. \ref{F3}(d)-(f) & Fig. \ref{F4}(a)-(c) & Fig. \ref{F4}(d)-(f) & Fig. \ref{F4}(g)-(i) \\ \hline
Random walk (s) & $1.17\times10^3$ & $7.09\times10^2$ & $7.09\times10^2$ & $1.93\times10^2$ & $3.43\times10^3$ & $8.87\times10^2$ & $5.27\times10^3$ \\ \hline
SSDA PDE (s) & $1.03\times10^0$ & $2.41\times10^0$ & $1.94\times10^0$ & $1.93\times10^0$ & $1.79\times10^0$ & $1.73\times10^0$ & $1.78\times10^0$\\ \hline
\end{tabular} }
\end{center}
\caption{Comparison between the time taken to perform $10^4$ realisations of the random walk process and the time taken to obtain the numerical solution to the SSDA PDE. All simulations were performed on a 3.1 GHz Intel i7 Processor (7920HQ) in MATLAB R2018b.}
\label{T1}
\end{table} 

In Table \ref{T1} we present a comparison of the computation time required to perform sufficient realisations of the random walk process to obtain representative average behaviour, and the time required to obtain a numerical solution to the SSDA PDE, for the results presented in Figures \ref{F2}-{F4}. For all cases, we observe that it is significantly faster, up to three orders of magnitude, to obtain predictions from the SSDA rather than the random walk.

\section{Discussion and conclusions}

Determining whether a population will undergo extinction is relevant throughout biology and ecology  \cite{anderson1992,axelrod2006,berger1990,keeling2007,neufeld2017,saltz1995,thomas2004}. Population extinction may be beneficial, such as in the case of treating a malignant tumour cell population \cite{coghlin2010}, or detrimental, such as in the case of endangered animal populations \cite{berger1990,saltz1995}. However, in both cases, being able to accurately predict whether population extinct will occur is invaluable. Lattice-based random walks with finite size effects are widely used to model the dynamics of populations of individuals \cite{bubba2019,fadai2019,johnston2014,johnston2014b}. Performing sufficiently many realisations of such random walks to obtain representative average behaviour is computationally intensive, which has motivated the use of continuum approximations of random walks \cite{baker2010}. However, previous continuum approximations are incapable of predicting population extinction. \\

The new continuum approximation of a lattice-based birth-death-movement random walk presented here is capable of describing and predicting population extinction. This approximation involves a state space representation of the lattice-based random walk, obtained via a mean-field assumption, followed by a continuum scaling of both the state space and time. Previous continuum approximations of lattice-based random walks have focused on the average occupancy of the lattice, which is ill-suited for analysing extinction \cite{baker2010,simpson2011}. In contrast, the SSDA explicitly accounts for the possibility that an individual random walk will undergo extinction via the distribution of the number of lattice sites occupied by individuals. \\

We have demonstrated that the SSDA faithfully captures the behaviour in the corresponding random walk for two common models of population dynamics, the logistic growth model and the Allee growth model \cite{baker2010,fadai2019,johnston2017}. Specifically, we have shown that the evolution of average occupancy obtained from the SSDA matches the average occupancy obtained from the corresponding random walk. This is in contrast with the standard mean-field ODE, which provides inaccurate predictions of the evolution of average occupancy. Further, the SSDA provides further relevant information about the average population behaviour, namely, the occupancy distribution and the probability that the population has undergone extinction. Neither of these measures can be obtained from the standard mean-field ODE. We have demonstrated that this additional information accurately reflects the behaviour in the corresponding random walk. Finally, we have highlighted the benefit of using the SSDA rather than the random walk by comparing the computational time required to obtain representative population behaviour. \\

Significant research effort has been directed toward continuum approximations of lattice-based random walks \cite{fadai2019,codling2008,baker2010,simpson2011,johnston2016b,markham2013}. In particular, the vast majority of previous work has investigated the validity of the standard mean-field assumption for a suite of biologically-inspired mechanisms \cite{baker2010,simpson2011,johnston2016b,markham2013}. It is well known that the mean-field assumption is satisfied provided that the positions of individuals in the population are not correlated \cite{baker2010,sharkey2008}. That is, the probability that a particular lattice site is occupied does not depend on the occupancy of other lattice sites. However, certain mechanisms in the random walk induce short-range correlation \cite{johnston2014b}. For example, the birth mechanism in the random walk considered here results in two individuals occupying neighbouring lattice sites. Hence, without sufficiently many movement events, compared to birth events, the mean-field assumption will be invalid due to this short range correlation. To address this issue, corrected mean-field models have been proposed \cite{baker2010,simpson2011,johnston2016b,markham2013,johnston2015}. Corrected mean-field models consider the evolution of pairs of lattice sites, and hence can account for correlation between lattice sites. These models, and similar approaches, extend the parameter regimes under which the random walk can be accurately approximated \cite{baker2010,simpson2011,johnston2016b,markham2013,johnston2015}. However, all of these models have focused on the issue of correlation in the random walk, and how to select an appropriate continuum approximation, based on the strength of correlation between lattice sites in the underlying random walk. \\

In contrast, the continuum approximation considered here addresses the issue of population extinction. In standard continuum approximations, the evolution of the occupancy of the lattice is described via a continuum equation \cite{baker2010,simpson2011,markham2013,johnston2015}. As such, the average occupancy will never be exactly zero, and will eventually approach the stable steady state in the continuum approximation equation. However, in the corresponding random walk, there is a non-zero probability, monotonically increasing in time, that the population has undergone extinction. As we have demonstrated in this work, this extinction probability significantly impacts the evolution of the average occupancy, even if the mean-field assumption is valid. Therefore, to accurately predict the average occupancy, the probability that the population undergoes extinction must be considered. We expect that a technique that incorporates \textit{both} the probability of population extinction and the correlation in lattice site occupancy will be necessary to accurately predict population behaviour for parameter suites where the mean-field assumption is not valid. However, we leave this investigation for future work. \\

The work presented can here can be extended in several directions. While we have only considered a population of individuals that is a single species, a natural extension would be to consider more than one species \cite{johnston2015}. Increasing the number of species allows for the description of more detailed individual behaviour, such as predation. However, the size of the state space increases exponentially with the number of species; an $n$-species model would have a state space of size $O(N^n)$. This quickly becomes intractable to analyse via the state space representation, necessitating the use of an appropriate continuum approximation \cite{sharkey2008}. The combination of the SSDA and a multispecies model would allow for the analysis of competition between species, and which species undergo extinction or survive. Another potential future direction would be to develop further mechanisms in the underlying random walk model, informed by experimental data. This approach would not only develop new partial differential equations, but may also provide insight into experimental mechanisms that were hitherto unclear \cite{fadai2019}. \\

\enlargethispage{20pt}

\noindent \textbf{Data Access} The code used to generate the results presented here can be found on Github at \\ \text{https://github.com/DrStuartJohnston/state-space-diffusion-approximation} \\

\noindent \textbf{Author contributions} STJ conceived the study. STJ, MJS and EJC designed the analysis and numerical experiments. STJ performed the analysis and numerical experiments. STJ wrote the manuscript. STJ, MJS and EJC edited the manuscript. All authors gave final approval for publication. \\

\noindent \textbf{Competing interests} The authors declare that they have no competing interests. \\

\noindent \textbf{Funding} This work was in part supported by the Australian Research Council Centre of Excellence in Convergent Bio-Nano Science and Technology (project no. CE140100036) and the Australian Research Council (DP170100474).
%\ack{Acknowledgements here.}

\section*{Appendix A}
\subsection*{Numerical solution of the SSDA PDE}
To obtain numerical solutions of the SSDA PDE, we first apply finite difference techniques to the spatial derivatives in the governing equation, using a central difference approximation \cite{press2007}. For all results we choose a node spacing of $\Delta s = 10^{-5}$. The node spacing is selected such that the numerical solution was insensitive to further reduction in the node spacing. For the temporal derivative, we implement the backwards Euler method \cite{press2007} with time step $\Delta t = T_{\text{End}}/1000$. The resulting tridiagonal system of algebraic equations is solved using the Thomas algorithm \cite{press2007}.
\subsection*{Numerical solution of the mean-field ODE}
To obtain numerical solutions of the mean-field ODE, we use MATLAB's inbuilt \textit{ode45} routine, which implements an adaptive time step Runge-Kutta method \cite{shampine1997}.

\bibliographystyle{RS}

\end{document}